\documentclass{article}

\usepackage{arxiv}

\usepackage[utf8]{inputenc} 
\usepackage[T1]{fontenc}    
\usepackage{hyperref}       
\usepackage{url}            
\usepackage{booktabs}       
\usepackage{amsfonts}       
\usepackage{nicefrac}       
\usepackage{microtype}      
\usepackage{graphicx}
\usepackage{epstopdf, epsfig}
\usepackage{subfigure}

\title{Deep Reinforcement Learning achieves flow control of the 2D {K}{\'a}rm{\'a}n Vortex Street}
\author{
  Jean Rabault\\
  Department of Mathematics\\
  University of Oslo\\
  \texttt{jean.rblt@gmail.com} \\
   \And
  Ulysse Reglade \\
  CEMEF, Mines ParisTech\\
  \texttt{ulysse.reglade@mines-paristech.fr} \\
   \And
  Nicolas Cerardi \\
  CEMEF, Mines ParisTech\\
  \texttt{nicolas.cerardi@mines-paristech.fr} \\
   \And
  Miroslav Kuchta \\
  Department of Mathematics\\
  University of Oslo\\
  \texttt{mirok@math.uio.no} \\
   \And
  Atle Jensen \\
  Department of Mathematics\\
  University of Oslo\\
  \texttt{atlej@math.uio.no} \\
}

\begin{document}
\maketitle

\begin{abstract}
The {K}{\'a}rm{\'a}n Vortex Street has been investigated for over a century and offers a reference case
  for investigation of flow stability and control of high dimensionality, non-linear systems. Active flow control, while of considerable
  interest from a theoretical point of view and for industrial applications, has remained inaccessible
  due to the difficulty in finding successful control strategies. Here we show that Deep Reinforcement
  Learning can achieve a stable active control of the {K}{\'a}rm{\'a}n vortex street behind a two-dimensional
  cylinder.
  Our results show that Deep Reinforcement Learning can be used to design active flow
  controls and is a promising tool to study high dimensionality, non-linear, time dependent
  dynamic systems present in a wide range of scientific problems.
\end{abstract}


\section{Introduction}

The flow past a circular cylinder is a famous, intensively studied problem in fluid mechanics. Above a
critical Reynolds number (which measures the relative importance of inertia and viscosity in a flow and
is written as $Re = U L / \nu$, where $U$ is the typical velocity in the flow, $L$ a typical length scale,
and $\nu$ the kinematic viscosity), the flow results in an organized vortex street that has been studied
since the beginning of the 20th century \cite{von1911mechanismus}. While the transition to unsteady flow
at a critical Reynolds number $Re_{cr} \approx 46$ has been identified as a supercritical Hopf bifurcation
\cite{provansal1987benard}, a number of complex phenomena, some of which are still under discussion, make
the situation complex as $Re$ is increased. The detailed mechanisms behind the instability of the time-dependent
vortex street have been the subject of controversies in the past \cite{kida1982stabilizing, saffman1982inviscid}
and are still an active field of research \cite{mowlavi_arratia_gallaire_2016, heil2017topological}, while attempts
to perform active flow control face considerable challenges. Flow control is made challenging due to the unsteady,
non-linear, high dimensionality behavior implied by the Navier-Stokes (NS) equations as soon as inertia becomes
important \cite{marquet_sipp_jacquin_2008, barbagallo_dergham_sipp_schmid_robinet_2012}. Among the approaches used to
attack this problem, the methods based on the adjoint formulation of the optimal control, e.g. \cite{li2003optimal},
require a potentially large number of solutions to the adjoint system (in particular if the problem at hand is ill-posed),
and are thus hard to scale. A different
class of approaches is based on building a lower-dimensional representation of the flow, e.g. by projecting the full model
onto a lower dimensional state space obtained for example by Proper Orthogonal Decomposition or Dynamic Mode Decomposition \cite{ravindran2000reduced}.
However, the dimensionality of the space typically grows with $Re$ making construction of the models expensive. Furthermore,
the models may not be robust to changes of $Re$ outside of the training/construction range \cite{brunton2015closed}. While
strategies based on such models obtain good results on specific applications \cite{7569070}, these methods are difficult to
apply to strongly nonlinear dynamic systems and are very specific of a given configuration \cite{brunton2015closed}.
As a consequence, a recent review of the field of active flow control calls to embrace the Artificial Intelligence (AI) and Machine Learning paradigm
to perform control of complex high-dimensional systems \cite{brunton2015closed}. This view is confirmed by recent experiments
of closed-loop control \cite{debien2016closed}, \cite{gautier2015closed}, \cite{benard2016turbulent} where AI techniques,
in particular genetic programming, were applied with success.

Artificial intelligence and the machine learning paradigm are made even more attractive by several recent highly
profiled successes of Deep Artificial Neural Networks (DANNs), such as attaining super-human performance at image
labeling \cite{lecun2015deep}, crushing the leading human player at the game of Go \cite{silver2017mastering}, or
achieving control of complex robots \cite{7989385}, which have shed the light on their ability to handle complex,
non-linear systems. In particular, the Deep Reinforcement Learning (DRL) methodology is a promising avenue for
the control of complex systems. This approach consists in letting the DANN interact with the system it should
control through 3 channels: an observation of the state of the system, a control imposed by the network on the
system, and a reward function measuring control performance. The choice of the reward function allows to direct
the efforts of the DANN towards solving a specific problem.

Here, we present the first successful use of DANNs trained through DRL to attack the problem of active flow
control in a 2D simulation of the flow around a cylinder at a moderate value of the Reynolds number, $Re=100$ ($Re$ is computed
based on the cylinder diameter, for more details see the Supplementary Information).
This is an illustration of the potential of Artificial Neural Networks trained through Deep Reinforcement Learning
for studying and controlling high dimensionality, non-linear systems.

\section{Active flow control in a simulation}


To demonstrate the ability of DANNs trained through DRL to perform active flow control, we set up a simple 2D
simulation of the non-dimensionalized flow around a cylinder as described by the incompressible NS equations at $Re=100$. The
configuration chosen is a copy of a classical benchmark used for the validation of numerical codes \cite{Schafer1996}.
In addition, two small jets of angular width $10^{\circ}$ are set on the sides of the cylinder and
inject fluid in the direction normal to the cylinder surface, following the control set up by the DANN. This implies
that the control relies on influencing the separation of the {K}{\'a}rm{\'a}n vortices, rather than direct
injection of momentum as could be obtained through propulsion. The jets are controlled
through their mass flow rates, respectively $Q_1$ and $Q_2$. We choose to use synthetic jets, meaning
that no net mass flow rate is injected in the flow, which translates to the constraint $Q_1 + Q_2 = 0$. More details are available in the
Supplementary Information.

The aim of the control strategy is guided by the reward function fed to the DRL during training. In the
present work, we want to minimize the drag $D$ through a reduction in the strength of the vortex shedding.
For this, we define the reward function $r$ from both the drag $D$ and the lift $L$, following:

\[
  r = \langle D \rangle_{T} - |\langle L\rangle_{T}|,
\]

\noindent where $\langle\bullet\rangle_{T}$ indicates the mean over one full vortex shedding cycle. In the figures, we
present the value of the drag coefficient, which is a normalized value of the drag $C_D = \frac{D}{\rho \bar{U}^2 R}$,
where $\bar{U}=2 U(0) / 3$ is the mean velocity magnitude, $\rho$ the volumetric mass density of the fluid, and $R$ the diameter
of the cylinder (for more details, see the Supplementary Information). Similarly, the mass flow rates of the jets are
normalized as $Q^{*}_i = Q_i / Q_c$, where $Q_c = \int_{-R}^{R}{\rho U(y) dy}$ is the
mass flow rate introduced by the inlet profile that intersects the cylinder diameter. Therefore, $Q^{*}_i$ indicates the
relative strength of the control jets, compared with the incoming flow.

\begin{figure}
\begin{center}
\includegraphics[width=.95\textwidth]{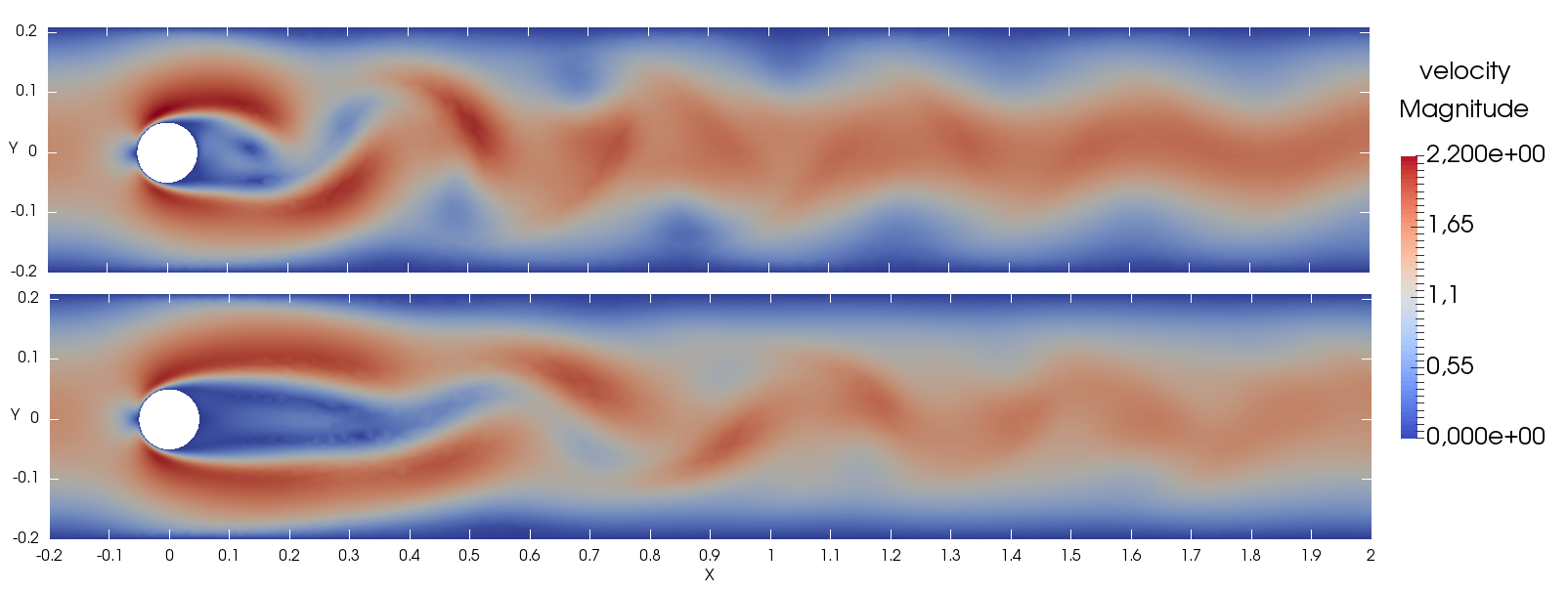}
\caption{\label{snapshot_comparison} Snapshots of the velocity magnitude illustrating the effect of the active
flow control. Top: baseline simulation. Bottom: result with active flow control. The size of the recirculation
bubble is greatly increased in the case with active control, and the strength of the {K}{\'a}rm{\'a}n vortex
street is reduced. A video comparing the two flows is available as Extended Data.}
\end{center}
\end{figure}

\begin{figure}
\begin{center}
\includegraphics[width=.70\textwidth]{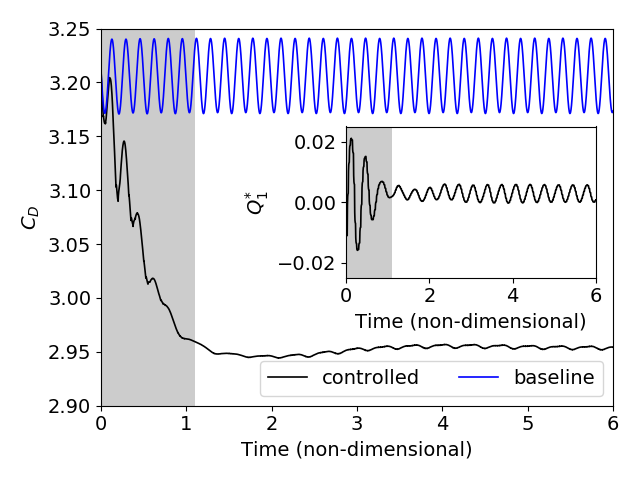}
\caption{\label{drag_control} Time-resolved value of the drag coefficient $C_D$ in the case without (baseline curve)
and with (controlled curve) active flow control, and corresponding normalized mass flow rate of the control jet 1 (inset).
The effect of the flow control on the drag is clearly visible: a reduction of the drag of around $8$\% is observed,
and the fluctuations in time due to vortex shedding are nearly suppressed. Two phases can be distinguished in the mass
flow rate control: first, a relatively large control is used to change the flow configuration, up to a non-dimensional
time of around $1.1$, before a pseudo periodic regime with very limited flow control is established.}
\end{center}
\end{figure}

Drastic changes are observed in the flow behind the cylinder in the case of active flow control by a trained
DANN compared with the baseline simulation without control, as visible in Fig. \ref{snapshot_comparison}.
A video of the flow undergoing active control, compared with the baseline, is also available as Extended Data.
The vortex shedding is both strongly reduced in intensity, and displaced around $3$ cylinder diameters downstream
of the cylinder. The change in the flow configuration has a notable effect on both the lift and drag: drag is
reduced by around $8$\% (see Fig. \ref{drag_control}), while the fluctuations in lift are reduced by around $75$\%.
The area $A$ of the recirculation bubble is drastically increased, by $125$\%. In addition, the vortex shedding
frequency is modified by the active control, and gets reduced by around $15$\%.

\begin{figure}
\begin{center}
\includegraphics[width=.95\textwidth]{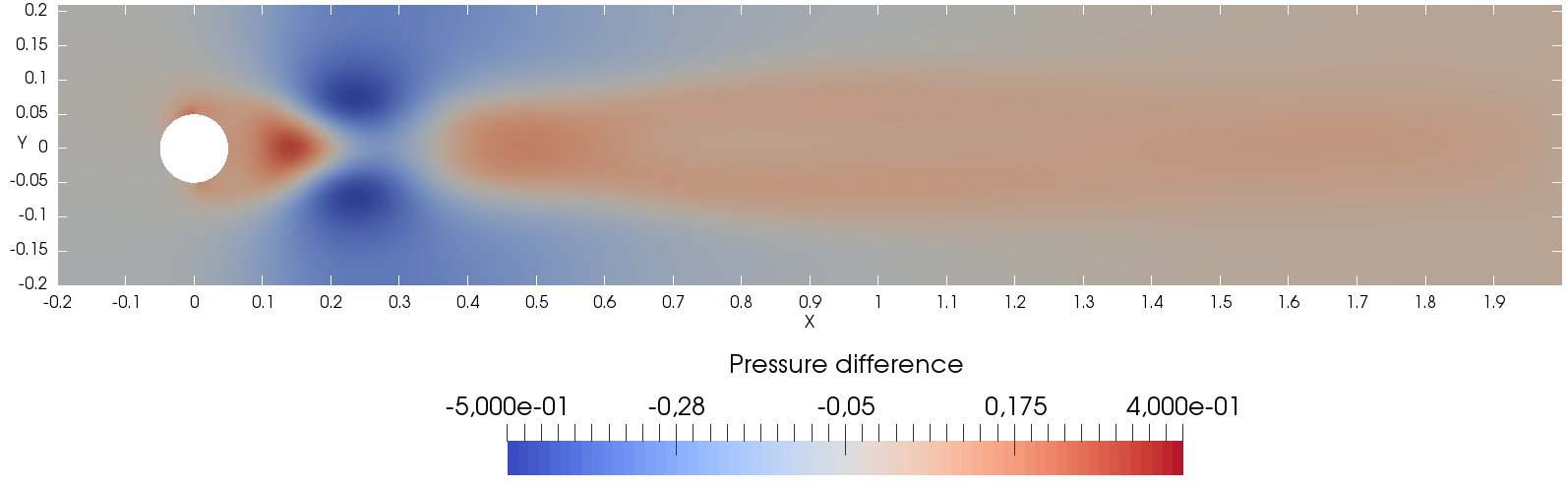}
\caption{\label{pressure_difference} Difference between the mean pressure field in the actuation case and the baseline case.
The pressure in the recirculation bubble is increased by the actuation, which is the cause for the drag reduction.}
\end{center}
\end{figure}

The explanation for the reduced drag is to be found in the change of the mean pressure distribution around the cylinder.
As visible in Fig. \ref{pressure_difference}, the recirculation area behind the cylinder undergoing active control is
both larger, and less intense than in the baseline case. This somewhat counter-intuitive effect is a good illustration
of the ability of the DANN to find complex strategies to optimize its reward function.

Finally, it should be noted that very little mass flow rate is necessary to perform active flow control, as visible
 in Fig. \ref{drag_control}. Two phases are visible on Fig. \ref{drag_control}, and also on the video in the
 Extended Data. First, a very short transient takes place where the DANN changes the flow configuration. Then,
 the DANN manages to keep the flow in this modified configuration, performing a pseudo-periodic control. While
 we give the DANN the freedom to set a control mass flow rate for each jet in the range of $Q^{*}_{i} = [-0.07, +0.07]$,
 in practice the network chooses to use normalized mass flow rates of absolute value at most $0.02$ during the initial
 transient control, and as small as $5 \times 10^{-3}$ at most to keep the vortex street in its new regime.
 This represents in average only around $0.35$\% of the incoming mass flow rate that intercepts the diameter of the cylinder
 for the active flow control in the established regime.
 Despite the very limited level of actuation in the new pseudo-periodic flow regime, altering the control
 signal even slightly leads to a collapse of the active control.

\section{Discussion and future work}

We prove that DANNs trained through DRL are a possible tool for performing active flow control. It is striking
 to see how efficient the DANN can be, as we show that our trained network uses very small controls to
 play with the state of a complex system. From the shape of the control signal, it seems that the DANN finds ways
 to control the flow through infinitesimal modifications of exponentially amplified modes that control the
 detachment of the vortex street, a fact that was previously reported in attempts to perform optimal flow control on simplified
 reduced order models \cite{7569070}.

Following our results, it appears that DANNs trained through DRL could have a general interest in fluid mechanics.
 Indeed, we have proved their relevance for active flow control, and a number of interesting cases should be
 investigated, such as the control of 3D Large Eddy Simulations, or Direct Numerical Simulation of
 boundary layers. In addition, one could try to combine DANNs with more traditional analytical methods to investigate
 properties of the solutions implied by the NS equations. Moreover, besides fluid mechanics, our work suggests that the use of DANNs
 as an experimental tool to study complex systems where traditional analytical approaches are limited should
be investigated. In this sense, our results can be seen as one more confirmation of the main message expressed in the last years
by the community investigating DANNs: DANNs are a successful tool at studying
and handling complexity, which under its many forms is one of the major challenges modern science still fails
to fully apprehend.

\section{Acknowledgments}

We gratefully acknowledge funding by the Research Council of Norway through the grants 'DOFI' (grant number 280625)
and 'RigSpray' (grant number 421160). We are grateful
to Terje Kvernes and Lucy Karpen for considerable help setting up the computation infrastructure used in this project.

\bibliographystyle{jfm}
\bibliography{references}

\begin{thebibliography}{26}
\expandafter\ifx\csname natexlab\endcsname\relax\def\natexlab#1{#1}\fi
\def\au#1{#1} \def\ed#1{#1} \def\yr#1{#1}\def\at#1{#1}\def\jt#1{\textit{#1}}
  \def\bt#1{#1}\def\bvol#1{\textbf{#1}} \def\vol#1{#1} \def\pg#1{#1}
  \def\publ#1{#1}\def\arxiv#1{#1}\def\org#1{#1}\def\st#1{\textit{#1}}

\bibitem[Abadi {\em et~al.\/}(2016)Abadi, Barham, Chen, Chen, Davis, Dean,
  Devin, Ghemawat, Irving, Isard {\em et~al.\/}]{abadi2016tensorflow}
{\sc \au{Abadi, Mart{\'\i}n}, \au{Barham, Paul}, \au{Chen, Jianmin}, \au{Chen,
  Zhifeng}, \au{Davis, Andy}, \au{Dean, Jeffrey}, \au{Devin, Matthieu},
  \au{Ghemawat, Sanjay}, \au{Irving, Geoffrey}, \au{Isard, Michael} \&
  \au{others}} \yr{2016} Tensorflow: A system for large-scale machine learning.
   \bt{In {\em OSDI\/}}, ,  \vol{vol.~16},  \pg{pp. 265--283}.

\bibitem[Atam {\em et~al.\/}(2017)Atam, Mathelin \& Cordier]{7569070}
{\sc \au{Atam, E.}, \au{Mathelin, L.} \& \au{Cordier, L.}} \yr{2017}
  \at{Identification-based closed-loop control strategies for a cylinder wake
  flow}.  \jt{IEEE Transactions on Control Systems Technology}  \bvol{25}~(4),
  \pg{1488--1495}.

\bibitem[Barbagallo {\em et~al.\/}(2012)Barbagallo, Dergham, Sipp, Schmid \&
  Robinet]{barbagallo_dergham_sipp_schmid_robinet_2012}
{\sc \au{Barbagallo, Alexandre}, \au{Dergham, Gregory}, \au{Sipp, Denis},
  \au{Schmid, Peter~J.} \& \au{Robinet, Jean-Christophe}} \yr{2012}
  \at{Closed-loop control of unsteadiness over a rounded backward-facing step}.
   \jt{Journal of Fluid Mechanics}  \bvol{703},  \pg{326–362}.

\bibitem[Benard {\em et~al.\/}(2016)Benard, Pons-Prats, Periaux, Bugeda, Braud,
  Bonnet \& Moreau]{benard2016turbulent}
{\sc \au{Benard, N}, \au{Pons-Prats, Jordi}, \au{Periaux, J}, \au{Bugeda, G},
  \au{Braud, P}, \au{Bonnet, JP} \& \au{Moreau, E}} \yr{2016}  \at{Turbulent
  separated shear flow control by surface plasma actuator: Experimental
  optimization by genetic algorithm approach}.  \jt{Experiments in Fluids}
  \bvol{57}~(2),  \pg{22}.

\bibitem[Brunton \& Noack(2015)]{brunton2015closed}
{\sc \au{Brunton, Steven~L} \& \au{Noack, Bernd~R}} \yr{2015}  \at{Closed-loop
  turbulence control: progress and challenges}.  \jt{Applied Mechanics Reviews}
   \bvol{67}~(5),  \pg{050801}.

\bibitem[Debien {\em et~al.\/}(2016)Debien, von Krbek, Mazellier, Duriez,
  Cordier, Noack, Abel \& Kourta]{debien2016closed}
{\sc \au{Debien, Antoine}, \au{von Krbek, Kai~AFF}, \au{Mazellier, Nicolas},
  \au{Duriez, Thomas}, \au{Cordier, Laurent}, \au{Noack, Bernd~R}, \au{Abel,
  Markus~W} \& \au{Kourta, Azeddine}} \yr{2016}  \at{Closed-loop separation
  control over a sharp edge ramp using genetic programming}.  \jt{Experiments
  in fluids}  \bvol{57}~(3),  \pg{40}.

\bibitem[Gautier {\em et~al.\/}(2015)Gautier, Aider, Duriez, Noack, Segond \&
  Abel]{gautier2015closed}
{\sc \au{Gautier, Nicolas}, \au{Aider, J-L}, \au{Duriez, Thomas}, \au{Noack,
  BR}, \au{Segond, Marc} \& \au{Abel, Markus}} \yr{2015}  \at{Closed-loop
  separation control using machine learning}.  \jt{Journal of Fluid Mechanics}
  \bvol{770},  \pg{442--457}.

\bibitem[Geuzaine \& Remacle(2009)]{geuzaine2009gmsh}
{\sc \au{Geuzaine, Christophe} \& \au{Remacle, Jean-Fran{\c{c}}ois}} \yr{2009}
  \at{Gmsh: A 3-{D} finite element mesh generator with built-in pre-and
  post-processing facilities}.  \jt{International journal for numerical methods
  in engineering}  \bvol{79}~(11),  \pg{1309--1331}.

\bibitem[Goda(1979)]{GODA197976}
{\sc \au{Goda, Katuhiko}} \yr{1979}  \at{A multistep technique with implicit
  difference schemes for calculating two- or three-dimensional cavity flows}.
  \jt{Journal of Computational Physics}  \bvol{30}~(1),  \pg{76 -- 95}.

\bibitem[Gu {\em et~al.\/}(2017)Gu, Holly, Lillicrap \& Levine]{7989385}
{\sc \au{Gu, S.}, \au{Holly, E.}, \au{Lillicrap, T.} \& \au{Levine, S.}}
  \yr{2017} Deep reinforcement learning for robotic manipulation with
  asynchronous off-policy updates.  \bt{In {\em 2017 IEEE International
  Conference on Robotics and Automation (ICRA)\/}},  \pg{pp. 3389--3396}.

\bibitem[Heil {\em et~al.\/}(2017)Heil, Rosso, Hazel \&
  Br{\o}ns]{heil2017topological}
{\sc \au{Heil, Matthias}, \au{Rosso, Jordan}, \au{Hazel, Andrew~L} \&
  \au{Br{\o}ns, Morten}} \yr{2017}  \at{Topological fluid mechanics of the
  formation of the {K}{\'a}rm{\'a}n-vortex street}.  \jt{Journal of Fluid
  Mechanics}  \bvol{812},  \pg{199--221}.

\bibitem[Kida(1982)]{kida1982stabilizing}
{\sc \au{Kida, Shigeo}} \yr{1982}  \at{Stabilizing effects of finite core on
  {K}{\'a}rm{\'a}n vortex street}.  \jt{Journal of Fluid Mechanics}
  \bvol{122},  \pg{487--504}.

\bibitem[LeCun {\em et~al.\/}(2015)LeCun, Bengio \& Hinton]{lecun2015deep}
{\sc \au{LeCun, Yann}, \au{Bengio, Yoshua} \& \au{Hinton, Geoffrey}} \yr{2015}
  \at{Deep learning}.  \jt{Nature}  \bvol{521}~(7553),  \pg{436}.

\bibitem[Li {\em et~al.\/}(2003)Li, Navon, Hussaini \& Le~Dimet]{li2003optimal}
{\sc \au{Li, Zhijin}, \au{Navon, IM}, \au{Hussaini, MY} \& \au{Le~Dimet, F-X}}
  \yr{2003}  \at{Optimal control of cylinder wakes via suction and blowing}.
  \jt{Computers \& Fluids}  \bvol{32}~(2),  \pg{149--171}.

\bibitem[Logg {\em et~al.\/}(2012)Logg, Mardal \& Wells]{logg2012automated}
{\sc \au{Logg, Anders}, \au{Mardal, Kent-Andre} \& \au{Wells, Garth}} \yr{2012}
  {\em Automated solution of differential equations by the finite element
  method: The FEniCS book\/}, ,  \vol{vol.~84}.  \publ{Springer Science \&
  Business Media}.

\bibitem[Marquet {\em et~al.\/}(2008)Marquet, Sipp \&
  Jacquin]{marquet_sipp_jacquin_2008}
{\sc \au{Marquet, Olivier}, \au{Sipp, Denis} \& \au{Jacquin, Laurent}}
  \yr{2008}  \at{Sensitivity analysis and passive control of cylinder flow}.
  \jt{Journal of Fluid Mechanics}  \bvol{615},  \pg{221–252}.

\bibitem[Mowlavi {\em et~al.\/}(2016)Mowlavi, Arratia \&
  Gallaire]{mowlavi_arratia_gallaire_2016}
{\sc \au{Mowlavi, Saviz}, \au{Arratia, Cristóbal} \& \au{Gallaire, François}}
  \yr{2016}  \at{Spatio-temporal stability of the {K}{\'a}rm{\'a}n vortex
  street and the effect of confinement}.  \jt{Journal of Fluid Mechanics}
  \bvol{795},  \pg{187–209}.

\bibitem[Provansal {\em et~al.\/}(1987)Provansal, Mathis \&
  Boyer]{provansal1987benard}
{\sc \au{Provansal, M}, \au{Mathis, C} \& \au{Boyer, L}} \yr{1987}
  \at{B{\'e}nard-von {K}{\'a}rm{\'a}n instability: transient and forced
  regimes}.  \jt{Journal of Fluid Mechanics}  \bvol{182},  \pg{1--22}.

\bibitem[Ravindran(2000)]{ravindran2000reduced}
{\sc \au{Ravindran, Sivaguru~S}} \yr{2000}  \at{A reduced-order approach for
  optimal control of fluids using proper orthogonal decomposition}.
  \jt{International journal for numerical methods in fluids}  \bvol{34}~(5),
  \pg{425--448}.

\bibitem[Saffman \& Schatzman(1982)]{saffman1982inviscid}
{\sc \au{Saffman, PG} \& \au{Schatzman, JC}} \yr{1982}  \at{An inviscid model
  for the vortex-street wake}.  \jt{Journal of Fluid Mechanics}  \bvol{122},
  \pg{467--486}.

\bibitem[Schaarschmidt {\em et~al.\/}(2017)Schaarschmidt, Kuhnle \&
  Fricke]{schaarschmidt2017tensorforce}
{\sc \au{Schaarschmidt, Michael}, \au{Kuhnle, Alexander} \& \au{Fricke, Kai}}
  \yr{2017} Tensorforce: A {T}ensorflow library for applied reinforcement
  learning. Web page.

\bibitem[Sch{\"a}fer {\em et~al.\/}(1996)Sch{\"a}fer, Turek, Durst, Krause \&
  Rannacher]{Schafer1996}
{\sc \au{Sch{\"a}fer, M.}, \au{Turek, S.}, \au{Durst, F.}, \au{Krause, E.} \&
  \au{Rannacher, R.}} \yr{1996} {\em Benchmark Computations of Laminar Flow
  Around a Cylinder\/},  \pg{pp. 547--566}.  \publ{Wiesbaden: Vieweg+Teubner
  Verlag}.

\bibitem[Schulman {\em et~al.\/}(2017)Schulman, Wolski, Dhariwal, Radford \&
  Klimov]{schulman2017proximal}
{\sc \au{Schulman, John}, \au{Wolski, Filip}, \au{Dhariwal, Prafulla},
  \au{Radford, Alec} \& \au{Klimov, Oleg}} \yr{2017}  \at{Proximal policy
  optimization algorithms}.  \jt{arXiv preprint arXiv:1707.06347} .

\bibitem[Silver {\em et~al.\/}(2017)Silver, Schrittwieser, Simonyan,
  Antonoglou, Huang, Guez, Hubert, Baker, Lai, Bolton {\em
  et~al.\/}]{silver2017mastering}
{\sc \au{Silver, David}, \au{Schrittwieser, Julian}, \au{Simonyan, Karen},
  \au{Antonoglou, Ioannis}, \au{Huang, Aja}, \au{Guez, Arthur}, \au{Hubert,
  Thomas}, \au{Baker, Lucas}, \au{Lai, Matthew}, \au{Bolton, Adrian} \&
  \au{others}} \yr{2017}  \at{Mastering the game of {G}o without human
  knowledge}.  \jt{Nature}  \bvol{550}~(7676),  \pg{354}.

\bibitem[Valen-Sendstad {\em et~al.\/}(2012)Valen-Sendstad, Logg, Mardal,
  Narayanan \& Mortensen]{valen2012comparison}
{\sc \au{Valen-Sendstad, Kristian}, \au{Logg, Anders}, \au{Mardal, Kent-Andre},
  \au{Narayanan, Harish} \& \au{Mortensen, Mikael}} \yr{2012}  \at{A comparison
  of finite element schemes for the incompressible {N}avier--{S}tokes
  equations}.  \bt{In {\em Automated Solution of Differential Equations by the
  Finite Element Method\/}},  \pg{pp. 399--420}.  \publ{Springer}.

\bibitem[Von~K{\'a}rm{\'a}n(1911)]{von1911mechanismus}
{\sc \au{Von~K{\'a}rm{\'a}n, Th}} \yr{1911}  \at{{\"U}ber den mechanismus des
  widerstandes, den ein bewegter k{\"o}rper in einer fl{\"u}ssigkeit
  erf{\"a}hrt}.  \jt{Nachrichten von der Gesellschaft der Wissenschaften zu
  G{\"o}ttingen, Mathematisch-Physikalische Klasse}  \bvol{1911},
  \pg{509--517}.

\end{thebibliography}

\clearpage

\section{Supplementary Information}

Here we present some additional details regarding the simulation set-up, the DANN, and the DRL algorithm used.
All the code used for both the simulation and the DANN trained through DRL is based on open source packages
(respectively, FEniCS \cite{logg2012automated}, Tensorforce \cite{abadi2016tensorflow} and Tensorflow, \cite{schaarschmidt2017tensorforce}).
All code is made available on the Github repository of the corresponding author [will be released upon publication].

\subsection{Simulation}

The 2D simulation environment is non-dimensionalized. Following the baseline \cite{Schafer1996},
it consists of a cylinder of radius $R = 0.05$ immersed in a box of total length $L=2.2$ along the X-axis
and height $H=0.41$ along the Y-axis. The kinematic viscosity is set as $\nu = 10^{-3}$, the fluid volumetric
mass density is $\rho = 1$. The origin of the coordinate system is at the center of the cylinder. The inflow profile
(set on the left wall) is parabolic, and follows the formula (cf. 2D-2 test case in \cite{Schafer1996}):

\[
  U(y) = 4 U_{m} (H/2 - y)(H/2 + y) / H^2,
  \
 \]

\noindent where $(U(y), V(y)=0)$ is the velocity vector. We follow the baseline in choosing $U_m=1.5$.
A no slip boundary condition is imposed on the top and bottom walls and on the solid walls of the cylinder.
The outflow boundary condition is imposed on the right wall of the domain. The Reynolds number, based on the mean
velocity magnitude $\bar{U}=2 U(0) / 3$, is $Re = 2 R \bar{U} / \nu = 100$. The domain is
discretized using Gmsh \cite{geuzaine2009gmsh}. With the mesh refined in the vicinity of the cylinder
the computational domain consists of a total number of $7142$ triangular elements. A non-dimensional,
constant numerical time step $dt = 5.10^{-4}$ is used. Moreover, two additional refined meshes, counting up to
$30000$ elements, were used for performing a mesh refinement study. The difference in drag between the
coarsest and the finest mesh is under $1$\%, therefore the coarsest mesh appears suitable for performing
our computations. In the case of the finest mesh, the time step had to be reduced to $dt = 3.10^{-4}$ to avoid
violating the CFL conditions and preventing the simulation from diverging.

In the interest of short solution time, the governing Navier-Stokes
equations are solved in a segregated manner \cite{valen2012comparison}. More precisely, the
Incremental Pressure Correction Scheme (IPCS method, \cite{GODA197976}) with an explicit treatment
of the nonlinear term is used. Spatial discretization then relies on the finite element method implemented
within the FEniCS framework \cite{logg2012automated}.

As stated in the main text, the jets are normal to the cylinder wall and are implemented on the sides of the
cylinder, at angles $\theta_1 = 90~^\circ$ and $\theta_2 = 270~^\circ$ relative to the flow
direction. The jets are controlled through their mass flow rate, $Q_i$, $i=1, 2$, and are set
through a parabolic velocity profile going to zero at the edges of the jet. The jet width is
set to $10~^\circ$.

Finally, information is extracted from the simulation and provided to the DRL agent. $151$
pressure probes are located in the vicinity of the cylinder and in its wake. The total drag $D$ on the
cylinder $C$ is computed following:
\[
D = \int_{C}(\sigma \cdot n)\cdot e_x\,\mathrm{d}S,
\]
where $\sigma$ is the Cauchy stress tensor, $n$ is the outer unit normal vector of the
cylinder surface and $e_x=(1, 0)$.

A benchmark of the simulation was performed by observing the drag value, the pressure fluctuations
and the Strouhal number $St = f L / \bar{U}$, where $f$ is the vortex shedding frequency. Results are in
good agreement with previously published data \cite{Schafer1996}. For example, we observe a mean absolute
value of the non-dimensional drag without actuation of $D = 0.16$, which corresponds to a drag coefficient
$C_D = \frac{D}{\rho \bar{U}^2 R} \approx 3.2$, in excellent agreement with the results reported by
\cite{Schafer1996} who obtain, depending on the solver used, a mean drag coefficient of typically $3.19$ to $3.22$.

\subsection{Artificial Neural Network and Deep Reinforcement Learning algorithm}

The DANN is a simple fully connected network, featuring one input layer, two consecutive fully connected hidden
layers of size $512$ neurons each, and one output layer. The classical rectified linear unit (ReLU) of positive
slope $1$ is used as an activation function. The input layer is connected to the $151$ probes immersed in the
simulation that measure the value of the flow velocity in the vicinity of the cylinder. The output layer sets
the mass flow rates of the jets. The DRL algorithm used for training, known in the literature as the
Proximal Policy Optimization method (PPO \cite{schulman2017proximal}), is the current state-of-the-art for
training DANNs to perform continuous control. Each training epoch is started from a converged, well-defined
{K}{\'a}rm{\'a}n vortex street. An episode duration lasts around $8$ vortex shedding periods.

As stated in the main body of the article, we define the reward function $r$ from both the drag $D$ and the lift $L$, following:

\[
  r = \langle D \rangle_{T} - |\langle L\rangle_{T}|,
\]

\noindent where $\langle\bullet\rangle_{T}$ indicates the mean over one full vortex shedding cycle.

It is difficult for the PPO algorithm to learn the necessity to set time-correlated, continuous control signals.
This is a consequence of the PPO trying at first purely random controls. Therefore, we added two limitations to
the control. First, the control value provided by the network is kept constant for a duration of around $10$\%
of the vortex shedding period. Second, the control effectively set in the simulation is made continuous in time.
To this end, the control at each time step in the simulation is obtained for each jet as $c_{s+1} = c_{s} + \alpha (a - c_{s}) $,
where $c_s$ is the control of the jet considered at the previous numerical time step, $c_{s+1}$ is the new control,
$a$ is the action provided by the PPO agent for the current set of time steps, and $\alpha = 0.1$ is a numerical
parameter. In practice, the exact value of $\alpha$ has little to say over the performance of the control.

It takes around $120$ training epochs, lasting altogether for about one night on a standard desktop computer using $1$
CPU core, for the DANN to learn a reasonably converged strategy similar to what is shown in the main body of the text.

\section{Multimedia files}

A video showing the velocity magnitude obtained both in the baseline case with no control (top) and in the
case with active control (bottom) on the same color scale is provided here: \url{https://folk.uio.no/jeanra/Research/comparison_baseline_actuation.avi}. The actuation jets are quite visible
in the first frames of the video, corresponding to the initial phase when the DANN modifies the flow configuration.
Later on, the jets are so weak that they are barely visible. Nonetheless, the weak actuation is vital to maintain flow control.

\end{document}